\documentclass[9pt]{IEEEtran}
\usepackage{amsmath,amsfonts}
\usepackage{algorithmic}
\usepackage{algorithm}
\usepackage{array}
\usepackage[caption=false,font=normalsize,labelfont=sf,textfont=sf]{subfig}
\usepackage{textcomp}
\usepackage{braket}
\usepackage{stfloats}
\usepackage{url}
\usepackage{verbatim}
\usepackage{graphicx}
\usepackage{cite}
\usepackage{graphicx} 
\usepackage{amsmath}
\usepackage{braket}
\usepackage{textcomp}
\usepackage{cleveref}
\usepackage{fancyvrb}
\usepackage{setspace}

\usepackage{booktabs}
\usepackage{multirow}
\usepackage{float}
\usepackage{svg}

\usepackage{graphicx} 
\usepackage{pgfplots}
\usepackage{subcaption} 
\usepackage{amsmath}
\usepackage{tikz}
\usepackage{multirow}
\usepackage{multicol}
\usepackage{threeparttable} 
\newcommand*\circled[1]{\tikz[baseline=(char.base)]{
            \node[shape=circle,fill,inner sep=1.5pt] (char) {\textcolor{white}{#1}};}}

\usepackage{enumitem}
\usepackage{algorithmic}
\usepackage{array}
\usepackage{url}
\usepackage{setspace}

\AtBeginEnvironment{thebibliography}{\footnotesize}
\begin{document}

\title{HQNET: Harnessing Quantum Noise for Effective Training of Quantum Neural Networks in NISQ Era}

\author{\IEEEauthorblockN{Muhammad Kashif \IEEEauthorrefmark{1}\IEEEauthorrefmark{2}},
Muhammad Shafique\IEEEauthorrefmark{1}\IEEEauthorrefmark{2}

\IEEEauthorblockA{\IEEEauthorrefmark{1}  eBrain Lab, Division of Engineering, New York University Abu Dhabi, PO Box 129188, Abu Dhabi, UAE}\\
\IEEEauthorblockA{\IEEEauthorrefmark{2} \normalsize Center for Quantum and Topological Systems, NYUAD Research
Institute, New York University Abu Dhabi, UAE}

Emails: \{muhammadkashif, muhammad.shafique\}@nyu.edu

\vspace{-10pt}
}

\maketitle
\begin{abstract}

Effective training of Quantum Neural Networks (QNNs) is crucial in the Noisy Intermediate-Scale Quantum (NISQ) era, where noise accelerates the onset of barren plateaus (BPs) and limits scalability. This paper investigates how quantum noise impacts QNN trainability and demonstrates that careful selection of qubit measurement observables can mitigate these effects.
We analyze PauliX, PauliY, PauliZ, and a customized Hermitian observable under both global (all-qubit measured) and local (single-qubit measured) cost functions. Our results show that with global cost function, PauliX and PauliY lead to flatter landscapes under noise, while PauliZ maintains training up to $8$ qubits before encountering BPs. The customized Hermitian observable proves most robust, enabling training up to $10$ qubits in noisy settings. 
For local cost function setting, PauliZ outperforms PauliX and PauliY, maintaining efficiency up to $10$ qubits. These findings highlight the importance of noise-aware observable selection, offering a practical strategy to improve QNN performance and advance quantum machine learning in noisy environments.
\end{abstract}

\vspace{-10pt}

\begin{spacing}{0.96}
\section{Introduction}
The advent of Noisy Intermediate-Scale Quantum (NISQ) devices marks a significant step toward practical quantum computing \cite{Preskill:2018}. By harnessing superposition and entanglement, NISQ devices offer potential advantages in solving classically intractable problems in optimization and simulation \cite{liang:2023,fan:2023,kashif2025computational}. However, their utility is constrained by high noise levels and limited error correction, which become increasingly problematic with scale \cite{Preskill:2018,lau:2022,kashif2024investigating}. Despite these challenges, NISQ devices are already showing promise in cryptography \cite{renner:2023}, drug discovery \cite{pyrkov:2023}, and quantum machine learning \cite{Biamonte:2017,Benedetti_2019,innan:2025_next}. 

Variational Quantum Algorithms (VQAs) are a hybrid class of algorithms that leverage both quantum and classical computation, making them well-suited for NISQ devices \cite{cerezo:2021}. They solve optimization and simulation problems by using quantum circuits to encode solutions and classical optimizers to adjust circuit parameters.
Quantum Neural Networks (QNNs) are a subclass of VQAs designed to emulate classical neural networks using variational quantum circuits (VQCs) \cite{cerezo:2021,zaman2023survey}. These models are trained by tuning circuit parameters to perform tasks such as classification, regression, and pattern recognition \cite{Biamonte:2017,Farhi:2018,kashif:2021}. Despite their potential, QNNs face scalability issues due to Barren Plateaus (BPs). BPs are the regions in the cost function landscape where gradient variance vanishes exponentially with increasing qubit count \cite{McClean:2018}. This severely impedes optimization and poses a major barrier to the practical deployment of QNNs, necessitating focused research on mitigating BPs .

\begin{figure}[htbp]
    \centering
    \includegraphics[scale=0.55]{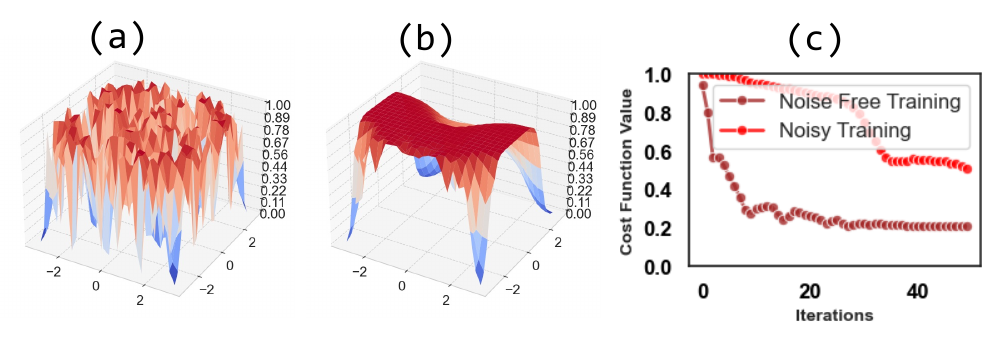}
    \caption{\footnotesize Cost Function Landscapes of 4-qubit QNN (a) Without Noise and (b) With Noise. (c) Comparison of Nosiy and Noise Free Training of 4-qubit QNN. The noise makes the cost function landscape flattened indicating the presence of BPs which eventually hinders the trainability of QNNs.}
    \label{fig:motiv_analysis}
\end{figure}

Several strategies have been proposed to mitigate BPs \cite{Liu_2023,Kashif:2023alleviating,kashif2025deep,kashif2024resqnets,kashif2024dilemma}, alongside efforts to understand their underlying causes. Key contributing factors include the nature and degree of entanglement \cite{Marrero:2020}, circuit expressibility \cite{kashif_unified}, cost function globality \cite{Cerezo:2021aa,Kashif:2023impact}, and hardware-induced noise \cite{ahmed2025quantum,Wang:2021}. Among these, quantum noise is particularly critical in the NISQ era, as it is inherent in current quantum devices and directly impacts the trainability of QNNs \cite{ahmed2025noisyhqnns,kashif2025nrqnn,ahmed2025quantum}. Addressing noise-related challenges is thus essential for advancing the practical deployment of QNNs.

\subsection{Motivational Example}
The robustness of QNNs to noise in NISQ devices remains a critical concern. To highlight how noise can give rise to early occurence of BPs leading to reduced trainability, Fig. \ref{fig:motiv_analysis} illustrates its impact on the cost function landscape. We employ a $4$-qubit QNN composed of alternating $R_x(\theta)$ and $R_y(\theta)$ rotations, with adjacent qubits entangled via controlled-Z (CZ) gates. The circuit depth is $10$, comprising $80$ single-qubit and 40 two-qubit gates, with measurements in the Pauli-Z basis. The model is trained to learn the identity function using the Adam optimizer (learning rate = $0.1$) for $50$ iterations.
Under ideal conditions, the cost landscape (Fig. \ref{fig:motiv_analysis}(a)) exhibits a rich structure and multiple minima, enabling efficient optimization. However, in noisy settings where each qubit is affected in every layer, the landscape flattens significantly (Fig. \ref{fig:motiv_analysis}(b)), impeding convergence. This degradation is evident in the training performance (Fig. \ref{fig:motiv_analysis}(c)), where noise-free training substantially outperforms the noisy counterpart.
These findings underscore the importance of addressing noise to realize scalable QNNs. This paper presents a detailed analysis of noise-induced effects on QNNs training, identifying conditions where noise may be leveraged and proposing strategies for its mitigation.


\subsection{Our Contributions}
We investigate how quantum noise and observable choice affect QNN training dynamics in global and local cost functions under fixed-depth circuits with increasing qubits. Our contributions are summarized below:


\begin{itemize}
    \item
    Our analysis reveals that quantum noise accelerates the onset of BPs in QNNs with global cost function definitions—an effect that typically occurs at higher qubit counts under noise-free conditions. However, we demonstrate that selecting appropriate qubit measurement observables, aligned with the cost function and target circuit output, can mitigate or even exploit this effect. To this end, we evaluate multiple measurement strategies, including PauliZ, PauliX, PauliY, and a custom-designed Hermitian observable.


    \item In the global cost function setting, we observe that the optimization landscapes for PauliX and PauliY measurements become increasingly flat as the number of qubits increases, which is a clear indication of BPs, and is further exacerbated by noise. In contrast, PauliZ measurement maintains performance up to $6$ qubits and shows moderate degradation at $8$ qubits under noisy conditions, with BPs becoming evident at $10$ qubits (maximum width considered in this paper).

    \item A notable finding is that QNNs utilizing the customized Hermitian observable with a global cost function exhibit noise-induced benefits, resulting in a truncated yet structured optimization landscape that facilitates convergence. This observable enables effective training up to $10$ qubits, highlighting that careful observable selection can harness noise to improve QNN trainability.

    \item 
    In contrast to the global cost function, the PauliZ observable under a local cost function demonstrates strong resilience to noise, enabling efficient training up to $10$ qubits and underscoring its suitability for noisy environments. Interestingly, PauliX and PauliY exhibit minimal improvement under the local cost setting, with consistently flat optimization landscapes that persist across both global and local definitions, severely limiting their trainability.

\end{itemize}

Our findings demonstrate that appropriate observable selection can not only mitigate the detrimental effects of noise but can also leverage the noise to enhance the trainability of QNNs.



\section{Preliminaries} \label{sec:background}

\subsection{Qubit Measurement and Observables}

\paragraph{PauliZ Observable} The PauliZ observable, denoted by $\sigma_z$ or $Z$, is one of the three fundamental Pauli matrices in quantum mechanics. It plays a crucial role in describing the state of a qubit. The PauliZ matrix is defined as:

\begin{equation}
    \sigma_z = Z = \begin{pmatrix} 1 & 0 \\ 0 & -1 \end{pmatrix}
\end{equation}

This $2 \times 2$ Hermitian matrix acts on the two-dimensional complex Hilbert space of a single qubit. In quantum mechanics, $\sigma_z$ corresponds to measurements along the Z-axis, distinguishing between the computational basis states $\ket{0}$ and $\ket{1}$. Its eigenvalues are $+1$ and $-1$, with $+1$ associated with $\ket{0}$ and $-1$ with $\ket{1}$. Thus, a measurement in the PauliZ basis reveals whether the qubit is aligned ($\ket{0}$) or anti-aligned ($\ket{1}$) with the Z-axis.


\paragraph{PauliX Observable} The PauliX observable, denoted by $\sigma_x$ or $X$, is the second of the three Pauli matrices and is defined as:

\begin{equation}
    \sigma_x = X = \begin{pmatrix} 0 & 1 \\ 1 & 0 \end{pmatrix}
\end{equation}

This observable corresponds to measurements along the X-axis in the Bloch sphere representation. When applied to a qubit, it acts as a bit-flip operator, exchanging the computational basis states $\ket{0}$ and $\ket{1}$. The PauliX gate is thus analogous to the classical NOT gate, making it fundamental for qubit state manipulation.
The eigenvalues of $\sigma_x$ are $+1$ and $-1$, but its eigenstates differ from the standard basis. Instead, they are the symmetric and anti-symmetric superposition states: 
\[
\ket{+} = \frac{\ket{0} + \ket{1}}{\sqrt{2}}, \quad \ket{-} = \frac{\ket{0} - \ket{1}}{\sqrt{2}}
\]
These eigenstates remain invariant under the action of $\sigma_x$, making them the stable states for measurements in the PauliX basis.


\paragraph{PauliY Observable}
The PauliY observable, denoted by the symbol 
$\sigma_y$ or $Y$, is the third Pauli matrix and is represented as:

$$\sigma_y = Y = \begin{pmatrix} 0 & -i\\ i & 0 \end{pmatrix}$$
In quantum mechanics, the PauliY observable corresponds to the measurement of spin or polarization along the Y-axis. When applied to a quantum state, it provides information about the phase relationship between the basis states $\ket{0}$ and $\ket{1}$. The eigenvalues of this matrix are $+1$ and $-1$, which are related to these phase-adjusted states. It is similar to the X-gate but also introduces a phase shift. The Y-gate, flips the state of a qubit from $\ket{0}$ to $i\ket{1}$ and $\ket{1}$ to $-i\ket{0}$. The eigenstates of the PauliY observable are different from the computational basis states ($\ket{0}$ and $\ket{1}$) and are instead specific superpositions of these states with complex coefficients. These states are stable under the action of the PauliY observable.


\paragraph{Customized Hermitian Observable}
A \textit{Hermitian observable} in quantum mechanics refers to any observable represented by a Hermitian operator, denoted by $\hat{A}$. A Hermitian operator is a linear operator on a Hilbert space that is equal to its own Hermitian conjugate (or adjoint), i.e.,

\begin{equation}
    \hat{A} = \hat{A}^{\dagger}
\end{equation}

This implies that for any two vectors $\ket{a}$ and $\ket{b}$ in the Hilbert space, the inner product satisfies:
\[
\bra{a} \hat{A} \ket{b} = \left( \bra{b} \hat{A} \ket{a} \right)^*
\]

In quantum mechanics, Hermitian operators are associated with physical observables such as position, momentum, spin, and energy. The Hermitian property guarantees that all eigenvalues of $\hat{A}$, which correspond to the possible outcomes of a measurement are real. Moreover, eigenvectors associated with distinct eigenvalues are orthogonal, a feature that underpins the probabilistic interpretation of quantum states and the predictability of measurement results.

This general framework encompasses well-known observables like the Pauli matrices, but also allows for the design of custom Hermitian observables tailored to specific tasks in quantum algorithms or quantum neural network training.

\begin{figure*}[htbp]
    \centering
    \includegraphics[scale=0.4]{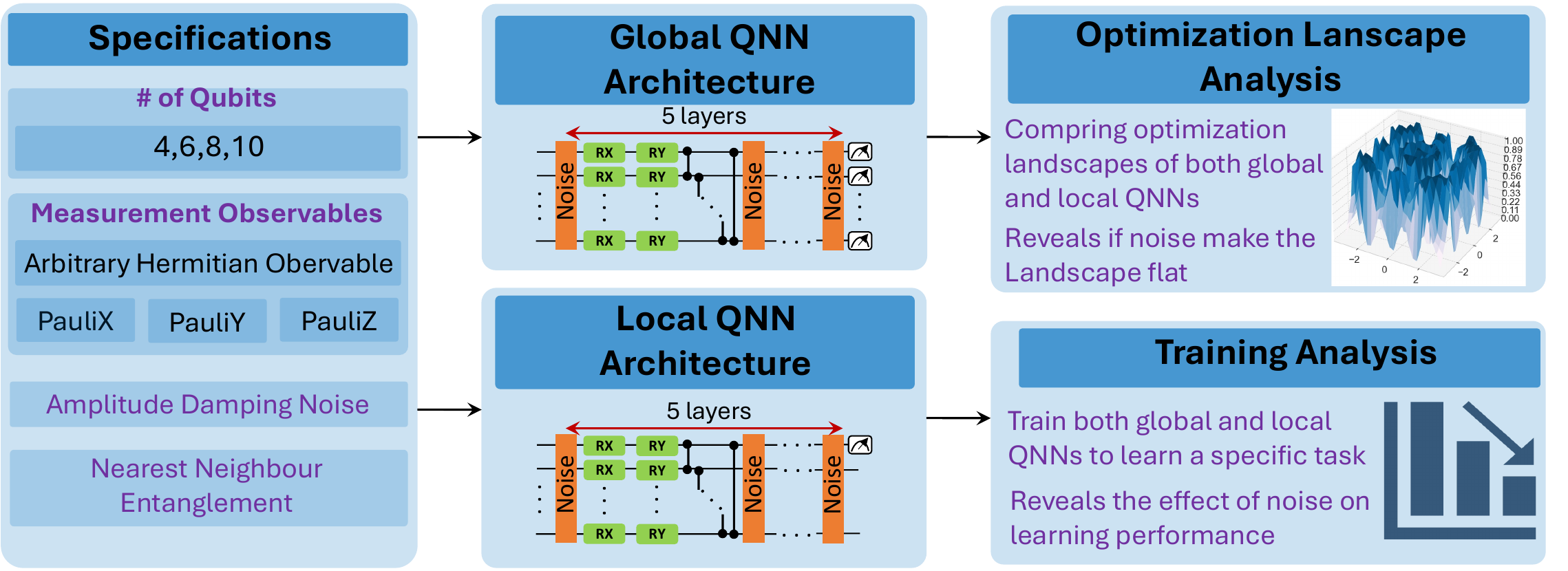}
    \caption{\footnotesize Detailed methodology highlighting key steps for the analysis of noise impact on the trainability of QNNs with different qubit measurement strategies. The quantum circuits used in QNN design are constructed with 4 to 10 qubits. Each qubit has RX and RY gates applied on it, and neighboring qubits are entangled via the CZ gate. Two distinct quantum circuit configurations were explored, Global QNN (all qubits were measured), and Local QNN (only a single qubit measured). The evaluation metrics used are optimization landscapes and cost function convergence. }
    \label{fig:methodology}
\end{figure*}
\subsection{Amplitude Damping}
Amplitude damping is a fundamental quantum noise model that describes energy dissipation in a quantum system, such as a qubit. It captures how interaction with the environment leads to energy loss, resulting in state transitions from excited to ground states. This process underlies physical phenomena such as spontaneous emission, scattering, and attenuation, and poses a significant challenge for reliable quantum computation.

The evolution of a quantum system under amplitude damping is commonly described using \textit{Kraus operators}, which characterize the dynamics of open quantum systems. The amplitude damping channel is represented by two Kraus matrices, $A_0$ and $A_1$, defined as:

\begin{enumerate}
    \item \textbf{$A_0$} — Describes the probability of the qubit remaining in its current state:
    \begin{equation}
        A_0 = \begin{pmatrix} 1 & 0\\ 0 & \sqrt{1-\gamma} \end{pmatrix}
    \end{equation}
    where $\gamma \in [0,1]$ is the damping parameter representing the probability of energy loss.
    
    \item \textbf{$A_1$} — Represents the transition of the qubit from the excited state to the ground state:
    \begin{equation}
        A_1 = \begin{pmatrix} 0 & \sqrt{\gamma}\\ 0 & 0 \end{pmatrix}
    \end{equation}
    This operator captures the decay of the qubit from $\ket{1}$ to $\ket{0}$ with probability $\gamma$ due to environmental interaction.
\end{enumerate}

Together, these Kraus operators satisfy the completeness relation $A_0^\dagger A_0 + A_1^\dagger A_1 = I$, ensuring that the quantum channel preserves trace and thus remains physically valid.








\section{HQNET Methodology}

We investigate the impact of quantum noise on the training dynamics of Quantum Neural Networks (QNNs), employing the hardware-efficient ansatz commonly used in NISQ-era applications. These circuits consist of parameterized single-qubit rotations and fixed two-qubit entangling gates, represented as:
\begin{equation}\label{eq1}
    U(\theta) = \prod_{i=1}^N U_{\text{ent}}U_{\text{rot}}(\theta_i)
\end{equation}
where $U_{\text{ent}}$ denotes the entangling gates, $U_{\text{rot}}(\theta_i)$ represents trainable single-qubit rotation gates, and $N$ is the number of circuit layers.
We conduct a comprehensive analysis to assess whether the quantum noise which is typically regarded as detrimental, can be strategically leveraged to improve optimization. Our methodology, illustrated in Fig.~\ref{fig:methodology}, systematically varies the number of qubits and measurement observables to uncover insights into the role of noise in QNN training.


\subsection{Specifications} \label{sec:specs}


\paragraph{Number of Qubits} 
An important aspect of our study involves examining the trainability of QNNs in relation to the phenomenon of BPs. To comprehensively assess this, we progressively increase the number of qubits. Our investigation starts with QNNs comprising 4 qubits, incrementally increasing in complexity up to QNNs with 10 qubits, i.e., $Q=\{4,6,8,10\}$. This stepwise increase helps in understanding that how the increase in qubit count influences the optimization landscape, particularly in the context of BPs and their impact on trainability under noise and noise-free environments.


\paragraph{Type of Observables} 
We used different observables for qubit measurement which are PauliZ, PauliX, PauliY and a customized hermitian observable. The details of these observables are already presented in Section \ref{sec:background}, however, hermitian operator that we have used is a specialized observable, constructed as a Hermitian matrix $H$, tailored to the dimensions of the quantum system under investigation. 
Typically, for a system of $n$ qubits, $H$ is represented by a $2^n \times 2^n$  matrix, initially populated with zeros in all its elements. A distinct modification is then introduced: the top-left element of $H$ denoted as $H_{0,0}$, is set to 1. This structure transforms $H$ into a high-dimensional analogue of the PauliZ matrix, extending its influence across the entire qubit ensemble. 
Mathematically, the Hermitian observable that we have used can be described as follows:

\[
H = \begin{bmatrix}
1 & 0 & \cdots & 0 \\
0 & 0 & \cdots & 0 \\
\vdots & \vdots & \ddots & \vdots \\
0 & 0 & \cdots & 0
\end{bmatrix}
\]
This corresponds to the projector $\ket{0000} \bra{0000}$, which projects onto the state where all qubits are in the $\ket{0}$ state.
This completely aligns with the training goal (cost function) considered in this paper, which is to learn an Identity gate (1 minus the  probability of all qubits being in $\ket{0}$ state), details of which are presented later in Section \ref{sec:exp_setup}.


\paragraph{Type of Noise - Amplitude Damping} 
Although, there can be different types of noise in NISQ devices which can effect the overall performance of a quantum algorithm. We considered the most common type of noise called as amplitude damping to analyze its impact on the learning performance of QNNs under consideration. 


\paragraph{Entanglement Type} 
Keeping in mind the limitations of NISQ devices, we used the QNNs with nearest neighbor entanglement, i.e., only the adjacent qubits are entangled.  


\subsection{QNN Architecture}
Once the required set of specifications are defined, we then contruct the QNNs for our analyis. We contruct two different QNNs differing mainly in the number of qubits measured. In the first QNN, which we call global QNN, we measure all the qubits in the underlying quantum layers. The second is called local QNN, where only a single qubit is measured. 
We use the hardware-efficient ansatz for quantum layers design, of the form as shown in Equation \ref{eq1}. 
Below we present the step-by-step details of our methodology:

\paragraph{Global QNN}
\begin{enumerate}
    \item The first step is to define and intialize the qubits. For every qubit number $\in Q$, the qubits are initialized on ground state:
    $$\ket{\psi_0} = \ket{0}^{\otimes n}$$

    \item Once the qubits are initialized, the next is to apply the unitary transformations. We apply two parameterized gates ($RX$ and $RY$) on each qubit:
    $$U_{rot} = \otimes_{i=1}^n RX{(\theta_i)} RY{(\phi_i)}$$

    where $\theta_i$ and $\phi_i$ are the rotation angles for the $i^{th}$ qubit.
    After the unitray operations, we also entangle the qubits, and as discussed preiously, we use the nearest neighbor entanglement:
    $$ U_{entangle} =  \prod_{i=1}^n CNOT_{i, i+1}$$

    where $CNOT_{i,i+1}$ entangle $i^{th}$ qubit with its neighbor.
    
    \item Given the above gate specifications, the QNN with a single layer is the form:
     $$ U =  U_{entangle} U_{rot}$$

    \item We consider the depth of QNN to be 5, i.e., the above mentioned single layer is repeated 5 times until measurement:
    $$\ket{\psi_{final}} = U^5 \ket{\psi_0}$$
    
    where the superscript 5 denotes the number of times the layer is repeated. Finally, the qubits are measured to get the output. For an customized Hermitian operator $M$, the probability of measuring a state $\ket{\phi}$ is given by:
        $$ P(\phi) = | \bra{\phi} M \ket{\psi_{final}}    |^2 $$

    The Pauli measurements (PauliX, PauliY and PauliZ) are similar to the Hermitian measurement but specifically with Pauli operators as discussed in Section \ref{sec:background}.

\end{enumerate}

\paragraph{Local QNN}
The architecture of local QNN is same as that of global QNN, which is discussed above, however, it differs in terms of qubit measurment, i.e., only a single qubit is measured. For instance, if the $i^{th}$ qubit is measured in PauliZ basis and the outcome is $0$, the post-measurement state of the system can be represented as:

$$ \ket{\psi'} = \frac{(I\otimes\ldots\otimes \bra{0}_i \otimes \ldots \otimes I) \ket{\psi_{final}}} {\sqrt{P(0)}}  $$

where, $P(0)$ is the probability of measuring the qubit in state $\ket{0}$, and 
$\ket{\psi_{final}}$ is the state of the system before measurement and the identity operators $I$ are applied to all qubits except the $i^{th}$ qubit.


\subsection{Optimization Landscape Analysis}
We conduct a detailed analysis of the optimization landscape of the QNNs under investigation. 
This involves plotting the cost function with respect to the network parameters, creating a landscape through which the optimizer traverses to identify the optimal solution. 
This analysis typically involves the identification and examination of local and global minima within these landscapes. Landscapes characterized by abundant and broader global minima are generally deemed conducive to effective optimization. 
Conversely, landscapes predominated by numerous local minima or extensive flat regions are typically considered less favorable for optimization purposes, presenting greater challenges in reaching the optimal solution.


\subsection{Training Analysis}
The QNNs with both local and global cost function definitions are then trained for particular problems. The training analysis encompasses a systematic examination of the convergence behavior of the cost function across a predetermined set of training iterations. This process involves scrutinizing how the cost function evolves and stabilizes during the training phase, thereby providing insights into the effectiveness and efficiency of the learning process over time.

\subsection{Experimental Setup} \label{sec:exp_setup}

The QNNs constructed in previous section are subjected to train for learning an Identity function. The cost function this context would be described by the following Equations. 


\begin{equation} \label{eq:CF}
    C = \bra{\psi(\theta)} (I-\ket{00\ldots0}\bra{00\ldots0}) \ket{\psi(\theta)} = 1- p_{\ket{00\ldots0}} 
\end{equation}


We typically want to maximize the probability of all the qubits being in $\ket{0}$ state for global cost function and the probability of first qubit in state $\ket{0}$ for local cost function. 
This objective closely aligns with the customized Hermitian observable that we have used (See Section \ref{sec:specs}) that projects the quantum system to an all $\ket{0}$ state in case of global cost function. 
For the training analysis, the QNNs are trained for $50$ training iterations. Adam optimizer with a learning rate of $0.1$ is used for the optimization. The experiments are performed using Pennylane \cite{Bergholm:2018}.   


\section{Results and Discussion}
\begin{figure*}[htbp]
    \centering
    \includegraphics[scale=0.41]{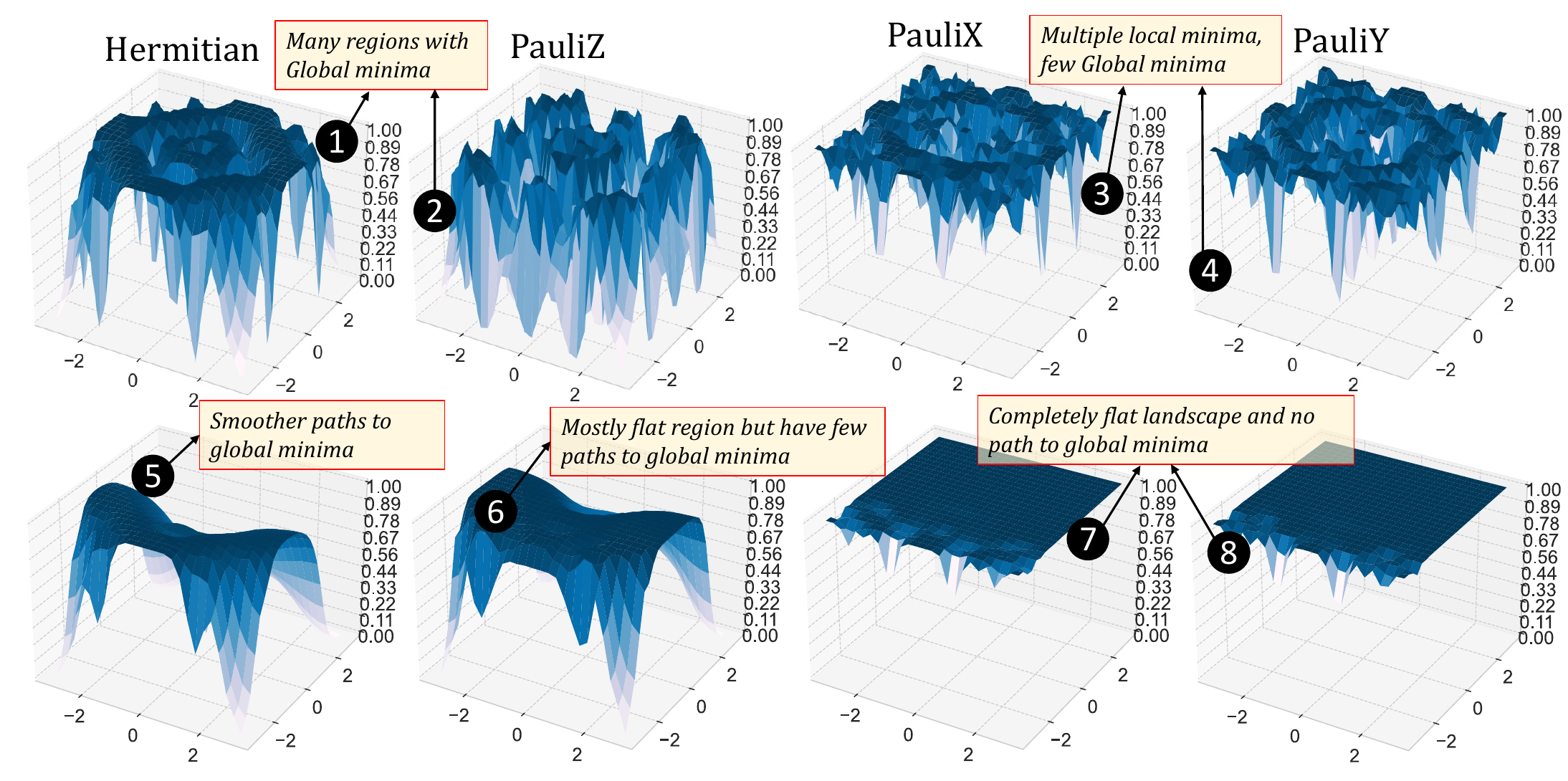}
    \caption{\footnotesize Optimization Landscape of 4-Qubit Global QNN under Noise-Free(upper panel) and Noisy (lower panel) Environments. For all the observables, the landscapes in case of no noise have multiple wider regions containing the solution whereas the landscapes with noise are mostly flat and truncated. For noisy environment, while using the customized Hermitian observable noise can be leveraged to advantage since the landscape trunction provides an easy path to solution.}
    \label{fig:4Q_LS}
\end{figure*}

\subsection{Global QNN - Comparative Analysis of QNNs Training Dynamics: Noisy vs. Ideal Conditions} \label{sec:results_global}
We first present a comparative analysis of the training dynamics of QNNs under both ideal (noise-free) and noisy conditions. 
This analysis focuses on 4 and 6-qubit QNNs, providing insights into the impact of noise on their training efficiency and optimization landscapes. 
This section aims to explain the distinct challenges and behaviors presented in the presence of noise, and how they differ from ideal, noiseless scenarios. 
This comparison is vital for understanding the resilience and adaptability of QNNs in real-world applications where noise is an inevitable factor.

\subsubsection{4-Qubit Global QNN}
\paragraph{Optimization Landscape in Noise-Free Setting}
In our investigation involving a 4-qubit quantum system, the overall architecture comprises 40 single-qubit gates and 15 two-qubit gates. The analysis of the cost function landscapes, as illustrated in Fig. \ref{fig:4Q_LS} (upper pannel), reveals distinct characteristics under ideal (noise-free) conditions. 
In such an environment, all landscapes demonstrate feasibility for optimization regardless of the measurement observable employed. However, a closer examination highlights that landscapes utilizing the customized Hermitian observable and PauliZ measurement observable are particularly more conducive to optimization. 
These landscapes exhibit multiple wider regions that converge to the solution as shown in label \circled{1} and \circled{2}, indicating a robust optimization potential. 
In contrast, landscapes associated with PauliX and PauliY measurement observables, while showing similarities to each other, are less optimal from the optimization perspective. They are characterized by limited presence of global minima and an increased presence of local minima, as shown in label \circled{3} and \circled{4} of Fig. \ref{fig:4Q_LS}. Despite this, they still maintain a moderate suitability for optimization, with a somewhat diminished, yet viable, efficiency in reaching the solution.

\begin{figure}[htbp]
    \centering
    \hspace{-20pt}
    \includegraphics[scale=0.45]{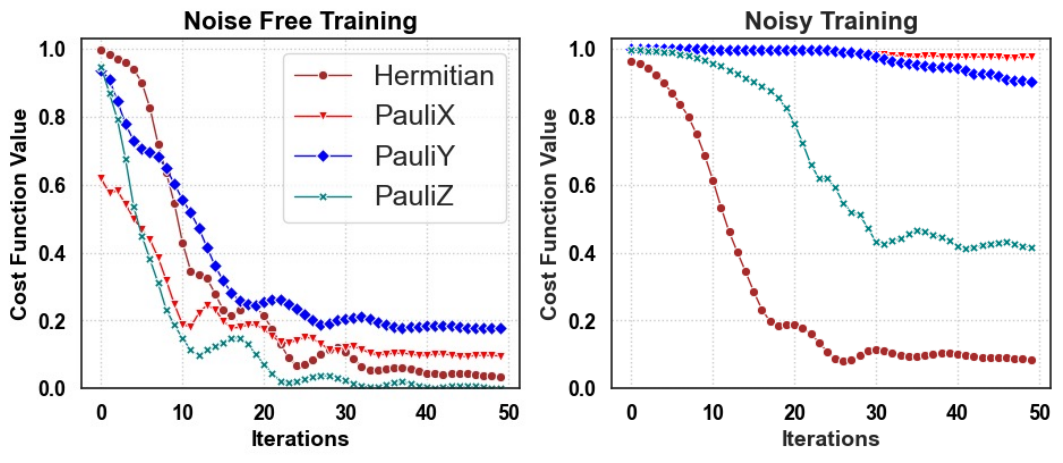}
    \caption{\footnotesize Training Results of 4-Qubit Global QNN Under Noisy and Noise-Free Environments. The noise-free training is better in all the cases, however, the customized Hermitian observable performs significantly better than other observables in noisy setting.}
    \label{fig:4Q_training}
\end{figure}

\begin{figure*}[htbp]
    \centering
    \includegraphics[scale=0.43]{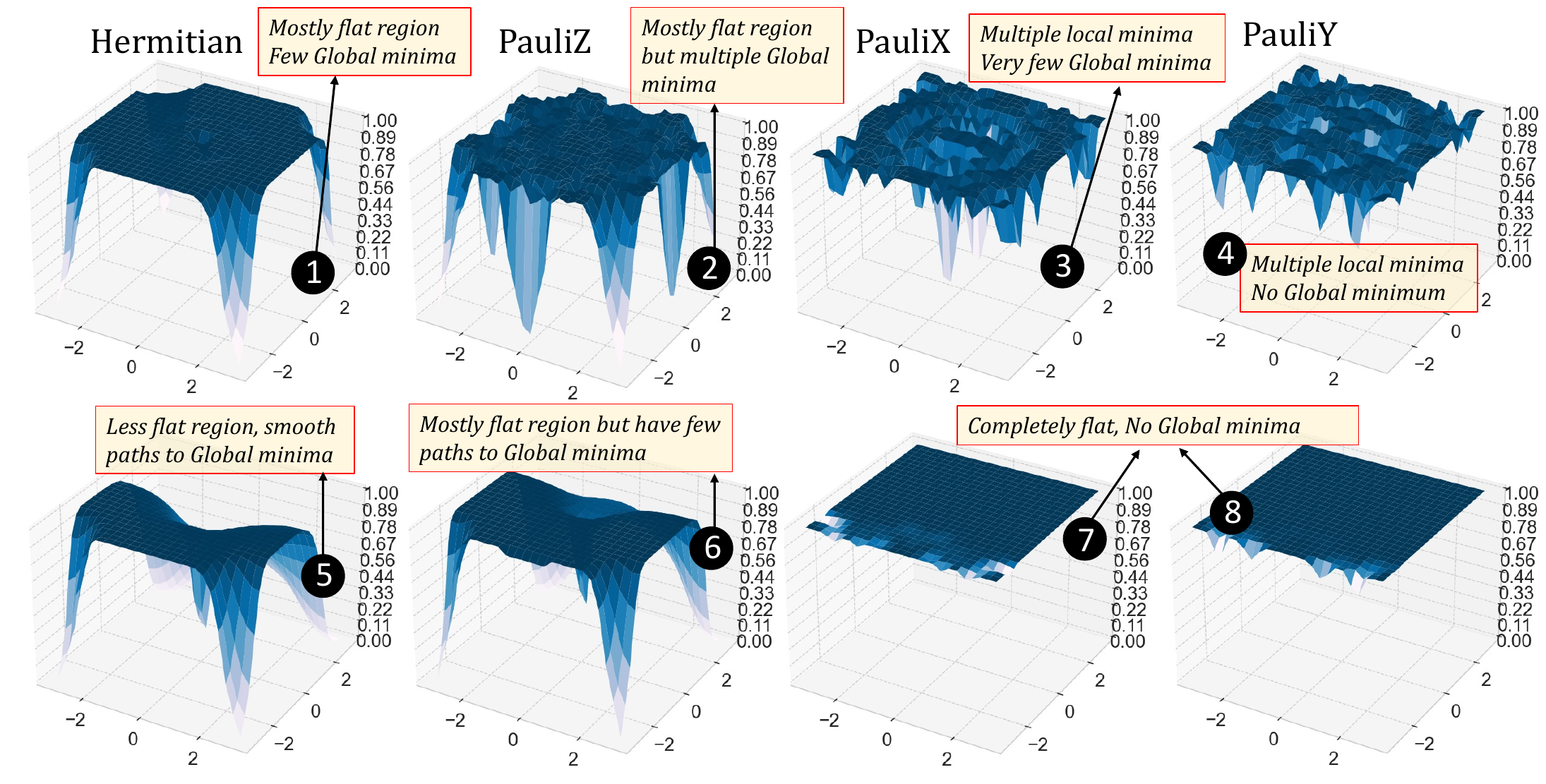}
    \caption{\footnotesize Optimization Landscapes of 6-Qubit Global QNN under Noise-Free (upper panel) Noise free) and Noisy (lower panel) Environments. 
    In case of no noise, the landscapes with the customized Hermitian and PauliZ observable seems better as they have few regions containing the global miminum, whereas with PauliX and PauliY observables there are multiple local minima and almost no global minimum. In the presence of noise, PauliX and PauliY observable's landscapes are completely flat with no room for optimization. The landscape with PauliZ observable is mostly flat with limited training potential. In case of the customized Hermitian observable noise can be leveraged to advantage since the landscape trunction provides an easy path to solution.
    }
    \label{fig:6Q_LS}
\end{figure*}

\paragraph{Optimization Landscape in Noisy Setting}
In scenarios where noise is introduced (Fig. \ref{fig:4Q_LS}(lower pannel), the optimization landscape dynamics alter significantly. The most resilient landscape under these conditions is observed with the Hermitian observable, where a minimal flat region and a broader, smoother trajectory towards the solution are observed, as shown in label \circled{5} in Fig. \ref{fig:4Q_LS}. 
The PauliZ observable's landscape, while exhibiting a slightly more flattened region compared to the Hermitian landscape, also maintains a broad and smooth trajectory leading to the solution (label \circled{6} in Fig. \ref{fig:4Q_LS}), and hence is also suitable to optimization. This indicates potential robustness of PauliZ observable against noise.

Conversely, the landscapes corresponding to PauliX and PauliY measurement observables undergo a substantial degradation in the presence of noise. 
They predominantly display flattened characteristics, with \emph{no} regions leading to the solution, as shown in label \circled{7} and \circled{8} in Fig. \ref{fig:4Q_LS}. This finding underscores an increased vulnerability to noise for these observables, common in NISQ devices, even in QNNs with shallow quantum layers. 
In contrast, landscapes associated with Hermitian and PauliZ observables display a significantly dynamic optimization landscape, exhibiting resilience to noise.


\paragraph{Training Comparison in Noisy and Noise-Free Settings}
Subsequent training of the 4-qubit QNN, for the problem defined in Equation \ref{eq:CF}, corroborates these landscape observations (Fig. \ref{fig:4Q_LS}), and the training results are depicted in Fig. \ref{fig:4Q_training}. 
Under noise-free conditions, training convergence is achieved across almost all measurement observables. However, in noisy environments, the Hermitian observable demonstrates superior performance with successful convergence. The PauliZ observable, while achieving a degree of training, exhibits suboptimal performance. PauliX and PauliY observables, aligning with their flat cost function landscapes, show negligible learning progress, reflecting their limited efficacy in noisy training scenarios, inevitable in NISQ devices.



\subsubsection{6-Qubit Global QNN}

\paragraph{Optimization Landscape in Noise-Free Setting}
In the 6-qubit QNN, comprising 60 single-qubit and 25 two-qubit gates, we observe distinct cost function landscapes for different measurement observables, as shown in Fig. \ref{fig:6Q_LS}(upper panel). 
Notably, the landscapes becomes more flatter compared to those in the 4-qubit system (Fig. \ref{fig:4Q_LS}), aligning with the BPs definition, which states that an increase in the number of qubits in QNNs leads to a vanishing gradient issue, resulting in progressively flatter landscapes. 
However, certain observables, notably Hermitian and PauliZ, demonstrate fewer local minima and multiple paths to the global minimum (label \circled{1} and \circled{2} in Fig. \ref{fig:6Q_LS}), offering effective optimization opportunities. Conversely, landscapes involving PauliZ and PauliY observables present multiple local minima with very few or no region containing the solution (label \circled{3} and \circled{4} in Fig. \ref{fig:6Q_LS}), posing a risk of trapping the optimizer in suboptimal solutions.

\begin{figure}[htbp]
    \centering
    \includegraphics[scale=0.36]{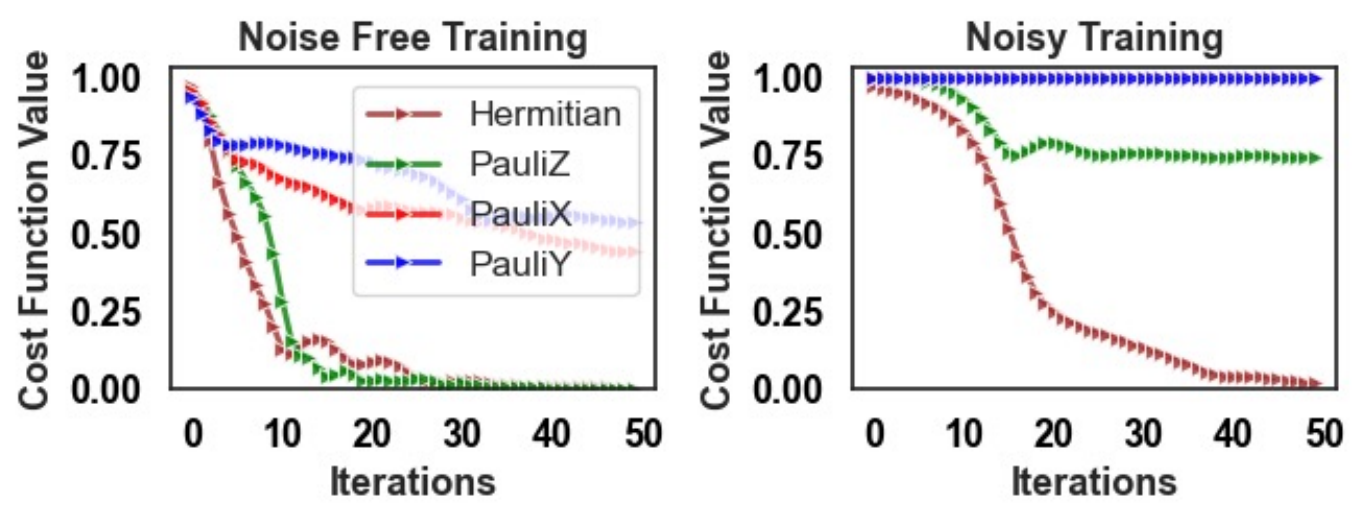}
    \caption{\footnotesize Training of 6-Qubit Global QNN Under Noisy and Noise-Free Environments. The noise-free training is better with all the observables than noisy training. However, the customized Hermitian observable performs significantly better than other observables and almosy same as in case noise-free training, in noisy setting.}
    \label{fig:6Q_training}
\end{figure}

\begin{figure*}[htbp]
    \centering
    \includegraphics[scale=0.41]{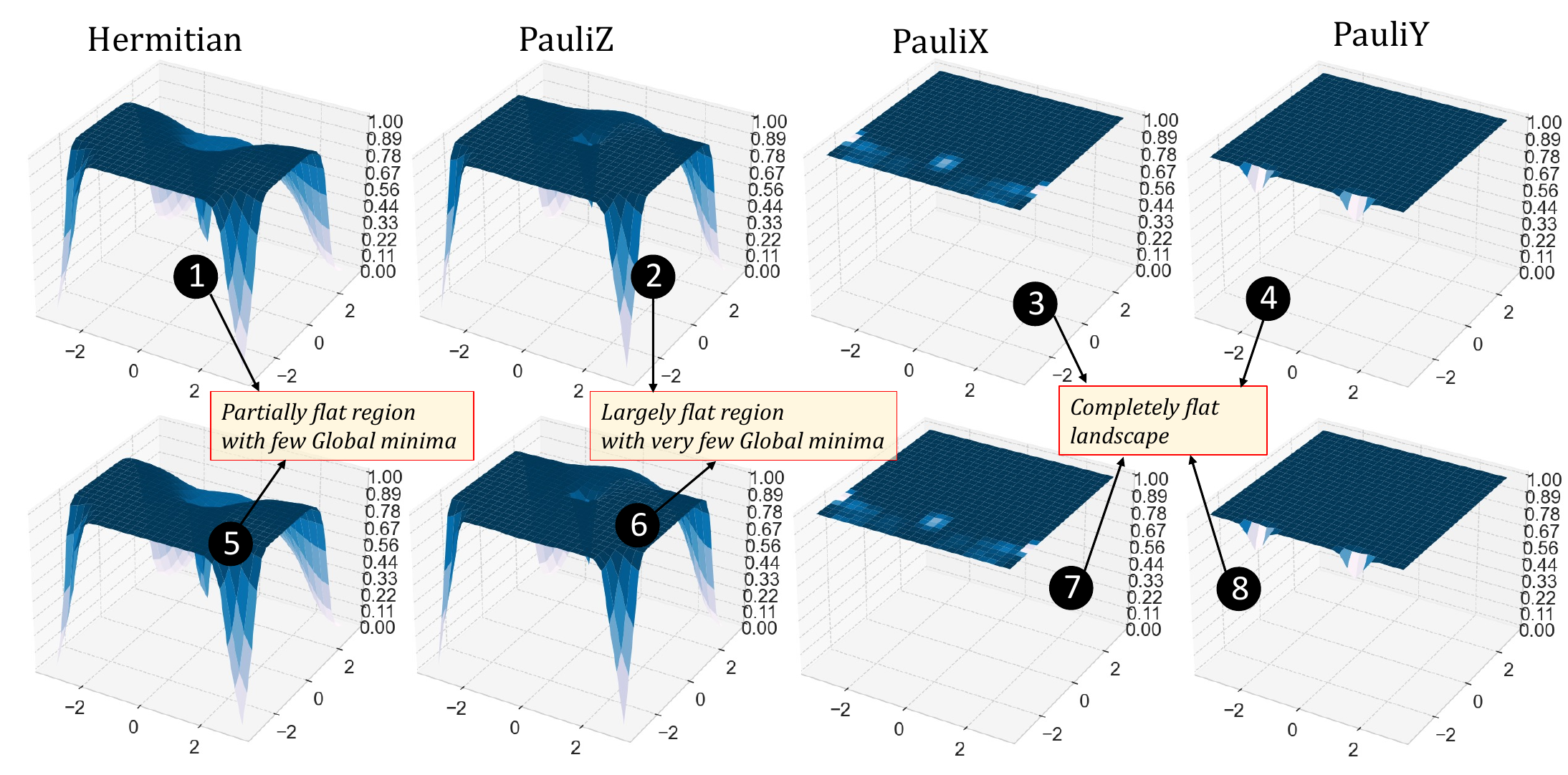}
    \caption{\footnotesize Optimization Landscapes under Noisy Setting for 8-Qubit (upper panel) and 10 Qubit(lower panel) Global QNN. The landscapes for QNNs with more expressive circuits tends to become further flat indicating the occurence of BPs. The landscape for PauliX and PauliY and PauliZ observables are completely and majorly flat with almost no or very limited potential for optimization. The landscape with customized Hermitian observable is starting to get partially flat but still provides a good smoother path to solution.}
    \label{fig:8_10Q_LS}
\end{figure*}

\paragraph{Optimization Landscape in Noisy Setting}
Introducing noise alters the optimization landscapes significantly as shown in Fig. \ref{fig:6Q_LS}(lower pannel). Optimization landscapes associated with PauliX and PauliY observables are the most affected, becoming completely flat offering no room for effective optimization, as shown in label \circled{7} and \circled{8} in Fig. \ref{fig:6Q_LS}. 
The PauliZ observable also shows a larger flat region in its optimization landscape and absence of the global minimum in the truncated optimization landscape and hence it will be difficult for the optimizer to navigate through to the solution (label \circled{6} in Fig. \ref{fig:6Q_LS}). 
Interestingly, the customized Hermitian observable's landscape, while truncated, becomes more conducive to optimization than in the noise-free setting (label \circled{1} in Fig. \ref{fig:6Q_LS}), with a reduced flat region and a wider path to the solution as highlighted in label \circled{5} of Fig. \ref{fig:6Q_LS}).

\paragraph{Training Comparison in Noisy and Noise-Free Settings}
The training efficacy of the 6-qubit system, as evaluated for the problem defined in Equation \ref{eq:CF} and illustrated in Fig. \ref{fig:6Q_training}, mirrors the observed landscape dynamics. In the absence of noise, most observables facilitate training, with Hermitian and PauliZ observables demonstrating superior performance, likely due to their landscape characteristics of minimal local minima and multiple global minima. In contrast, under noisy conditions, the Hermitian observable distinctly outperforms others, underscoring its robustness against noise interference.

Since we have already shown how the noise is affecting the peformance in QNNs, from this point onwards, we only present the optimization landscapes and training results for \emph{noisy} setting.


\subsection{Global QNN - Focused Analysis QNNs Training Dynamics in Noisy Setting}
Building on our established understanding of noise impact on QNN's performance, subsequent discussions will be centered exclusively on the noisy setting. 
This narrowed focus will shed light on the ways in which noise alters the optimization terrain in QNNs with more expressive quantum layers (8 and 10 qubits), with different qubit measurement strategies, offering insights into the challenges and potential strategies for achieving effective optimization under realistic, non-ideal conditions.


\subsubsection{8-Qubit Global QNN}

\paragraph{Analysis of Optimization Landscape}
The 8-qubit QNN consists of a total of 80 single and 35 two-qubit gates. The optimization landscapes of 8-qubit QNN with different measurement observables are shown in Fig. \ref{fig:8_10Q_LS} (upper pannel). We find trends consistent with our previous findings for 4 and 6-qubit QNNs. 
Specifically, the optimization landscapes for PauliX and PauliY observables are entirely flat, offering no opportunity for convergence to an optimal solution, as shown in label \circled{3} and \circled{4} in Fig. \ref{fig:8_10Q_LS}. 
Conversely, the landscape for PauliZ observables, though truncated due to noise, presents some potential for effective training, although with limited regions indicating the presence of a global minimum, as shown in label \circled{2} in Fig. \ref{fig:8_10Q_LS}, offering potential to somewhat mitigate the negative effects of BPs. 
Notably, the Hermitian observable demonstrates enhanced robustness in noisy conditions. The truncation in its landscape appears beneficial, providing a smoother and broader path towards the solution, as shown in label \circled{1} of Fig. \ref{fig:8_10Q_LS}, thus showing a significant potential to alleviate the adverse effects of BP even at higher qubit count.

\begin{figure}[htbp]
    \centering
    \includegraphics[scale=0.34]{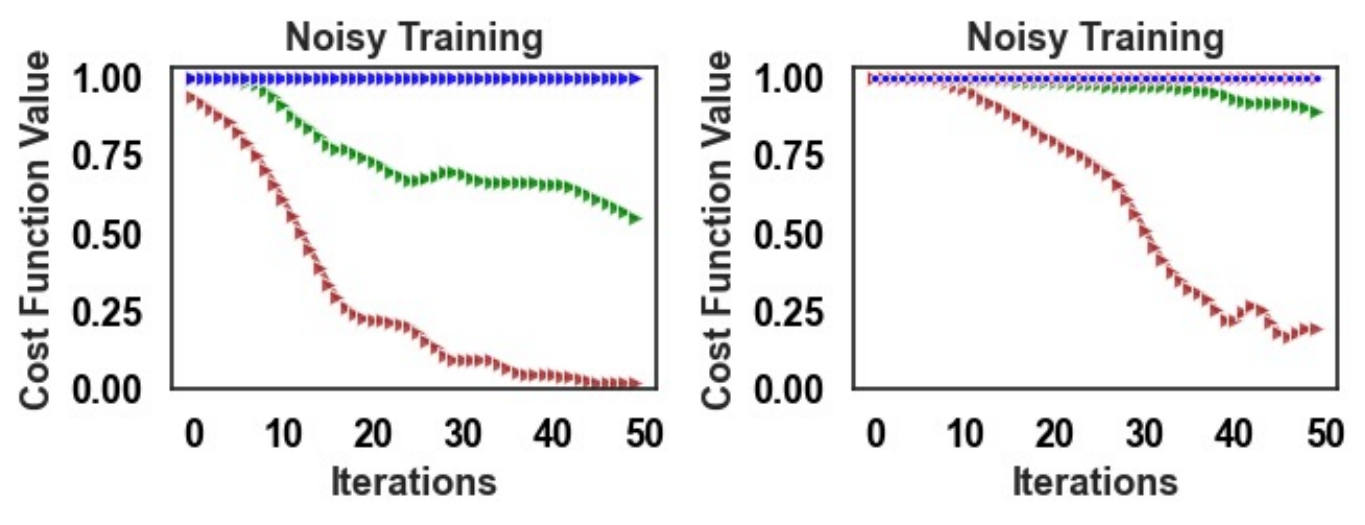}
    \caption{\footnotesize Training of 8 (left) and 10 (right) Qubit Global QNN with Different Observables Under Noise. For both 8 and 10 qubit QNNs, all the Pauli observables show limited or no training performance. Only the customized Hermitian observable performs effective training while benefitting from noise and escaping the BPs.}
    \label{fig:8_10Q_training}
\end{figure}

\paragraph{Training Analysis}
The training results of the 8-qubit circuit, as presented in Fig. \ref{fig:8_10Q_training}(left), align closely with the observed optimization landscapes. 
In line with the landscapes, the QNNs with PauliZ and PauliY observables exhibit negligible training progress. 
In contrast, QNNs using the PauliZ observable show moderate training effectiveness. 
Most impressively, QNNs employing the Hermitian observable demonstrate robust training performance, approaching optimal solutions and thereby highlighting the noise-resilience of the customized Hermitian observable used in this paper.

\subsubsection{10-Qubit Global QNN}
\paragraph{Analysis of Optimization Landscape.}
The 10-qubit QNN accounts to a total of 8
100 single and 45 two-qubit gates.
In Fig. \ref{fig:8_10Q_LS} (lower panel), we present the optimization landscapes associated with QNNs containing 10-qubit quantum layers under various measurement observables. It is observed that landscapes corresponding to PauliX and PauliY observables exhibit a completely flat topology, indicating an absence of optimization opportunities, as shown in label \circled{7} and \circled{8} in Fig. \ref{fig:8_10Q_LS}. 
Conversely, the landscape for the PauliZ observable, while predominantly flat, presents marginal optimization potential (label \circled{5} in Fig. \ref{fig:8_10Q_LS}). 
Notably, the landscape associated with an customized Hermitian observable is characterized by a truncated and less uniform topology (label \circled{5} in Fig. \ref{fig:8_10Q_LS}), suggesting a comparatively greater scope for optimization even in 10-qubit QNNs, potentially overcoming the BPs and enhancing the trainability potential of QNNs.

\paragraph{Training Analysis}
In Fig. \ref{fig:8_10Q_training}(right), we display the training outcomes for the 10-qubit circuit. Consistent with the dynamics observed in the optimization landscape, the model exhibits ineffective training when using any of the three Pauli observables (X, Y, Z). However, when employing the customized Hermitian observable, used in this paper, the model demonstrates effective training performance. This distinction underscores the enhanced suitability of the customized Hermitian observable (defined while keeping in mind the cost function and the output we want from QNN) for training more expressive quantum circuits.


\subsection{Local QNN: Analysis of QNNs Training Dynamics in Noisy Conditions}

In this section, we shift our focus to local QNNs, where only a single qubit is measured. 
Our previous investigations, detailed in Section \ref{sec:results_global}, demonstrated that a customized Hermitian observable, well-aligned with our chosen cost function (refer to Section \ref{sec:exp_setup}), outperformed standard observables (PauliZ, PauliX, and PauliY) in global QNN setup. Thus, this analysis will solely consider these standard observables to evaluate whether localizing the cost function enhances the overall performance of QNNs with these observables.
Furthermore, our previous results indicated that in the absence of quantum noise, all observables showed high effectiveness, especially in systems with lower qubit counts (4 and 6 qubits). However, under quantum noise and at higher qubit counts (8 and 10 qubits), these observables encountered barren plateaus (BPs) rapidly. Therefore, this analysis will concentrate on noisy conditions at these larger qubit counts.

\subsubsection{8-Qubit Local QNN}

\begin{figure}
    \hspace{-10pt}
    \includegraphics[scale=0.33]{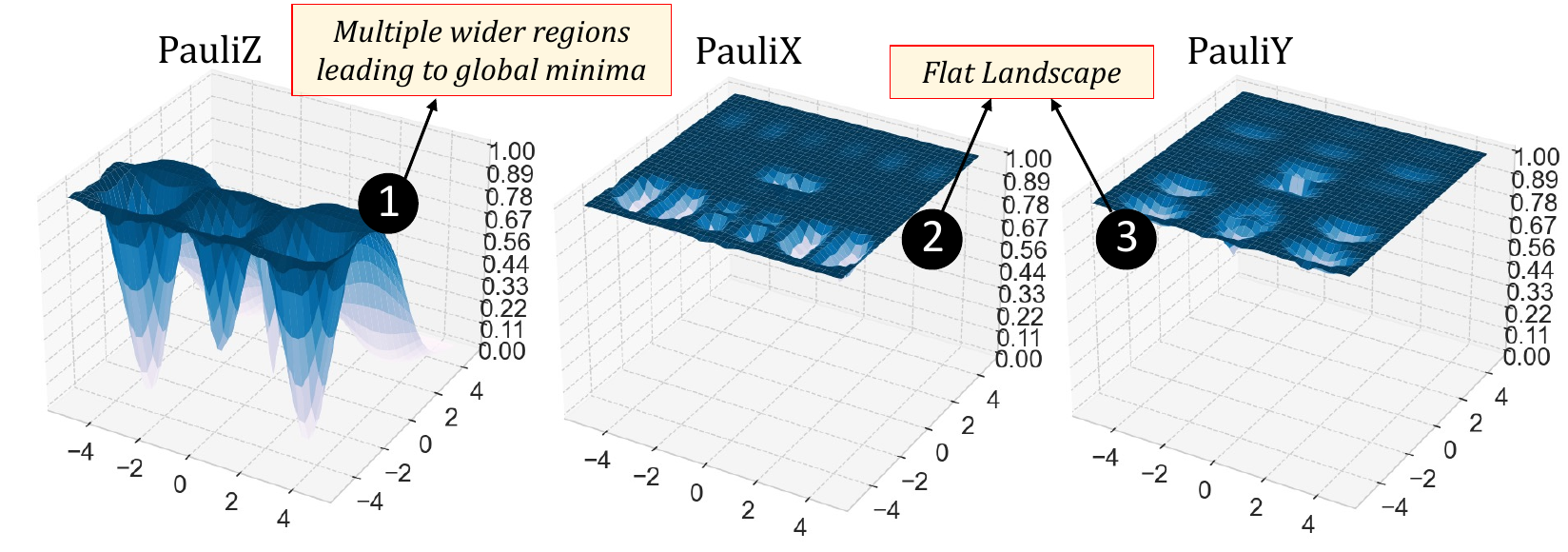}
    \caption{\footnotesize Optimization Landscape of 8-Qubit Local QNN under Noisy Settings with Different Observables. The landscapes with PauliZ observable in local QNN significantly benefits from noise than global QNN as there are multiple wider regions in its landscape leading to solution. PauliX and PauliY observables' landscape are still completely flat. }
    \label{fig:8qubit_localQNN_LS}
\end{figure}

The optimization landscape of an 8-qubit local QNN employing various observables is depicted in Fig. \ref{fig:8qubit_localQNN_LS}. The results reveals a notable similarity with the optimization landscape observed in global QNN configurations, as illustrated in Fig. \ref{fig:8_10Q_LS}(upper panel). 
Specifically, for the PauliX and PauliY observables, the optimization landscape predominantly exhibits flat regions, as shown in label \circled{2} and \circled{3} in Fig. \ref{fig:8qubit_localQNN_LS} indicating these observables are generally suboptimal for effective optimization irrespective of the cost fucntion globality and locality.
In contrast, the optimization landscape associated with the PauliZ observable in the local QNN setting demonstrates a significant divergence from its global QNN counterpart. In QNN with local cost function definition, the landscape for the PauliZ observable is not predominantly flat. Instead, it features multiple expansive regions that encompass the solution, as shown in label \circled{1} of Fig. \ref{fig:8qubit_localQNN_LS}. This characteristic suggests a more favorable scenario for optimization when using the PauliZ observable in QNNs with local cost function definition, in contrast to its behavior in QNNs with global cost function definition where a largely flat landscape is observed (label \circled{2} of Fig. \ref{fig:8_10Q_LS} (upper panel)).

This distinction in landscape topology between local and global QNN configurations, particularly with respect to the PauliZ observable, underscores the importance of choosing the appropriate observable in relation to the specific quantum layers architecture being utilized, to enhance the efficiency and success of the optimization process in QNNs.

\begin{figure}[htbp]
    \centering
    \includegraphics[scale=0.34]{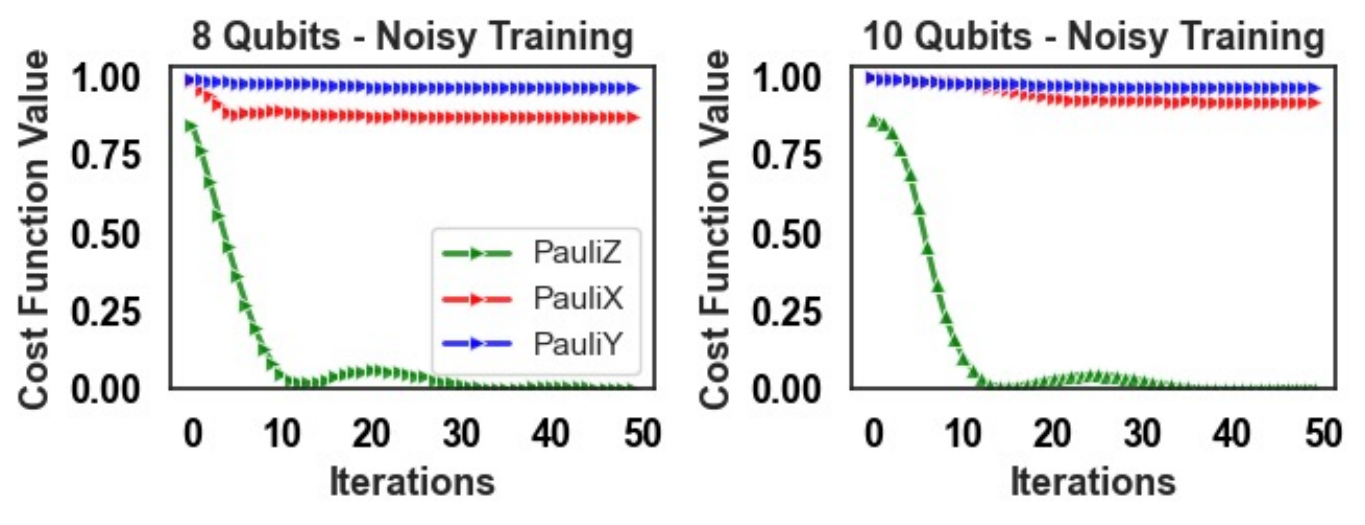}
    \caption{\footnotesize Training Results of $8$ and $10$ Qubit Local QNN with different observables in Noisy Environment. The landscape with PauliZ observable in local QNN even with $10$ qubits contains multiple wider regions leading to the solution. PauliX and PauliY observables' landscape are still completely flat.}
    \label{fig:8_10Q_training_local}
\end{figure}

The 8-qubit QNN with local cost function definition with different measurement observables are then subjected to training for the problem defined in Equation \ref{eq:CF}, the results of which are shown in Fig. \ref{fig:8_10Q_training_local}(left), which typically align with the optimization landscape results in Fig. \ref{fig:8qubit_localQNN_LS}(left).  
Specifically, the QNNs with PauliX and PauliY observables exhibit negligible or no training progress, in stark contrast to the QNN utilizing the PauliZ observable. Notably, the QNN with the PauliZ observable demonstrates significant training effectiveness, markedly surpassing the performance of the QNNs employing PauliX and PauliY observables.

This comparative analysis highlights the superior adaptability and learning efficiency of local QNN when the PauliZ observable is implemented, as opposed to the limited training efficacy observed with the PauliX and PauliY observables. Such outcomes underline the critical importance of selecting appropriate measurement observables in QNN configurations to achieve optimal training performance.


\subsubsection{10-Qubit Local QNN}
\begin{figure}[htbp]
    \includegraphics[scale=0.33]{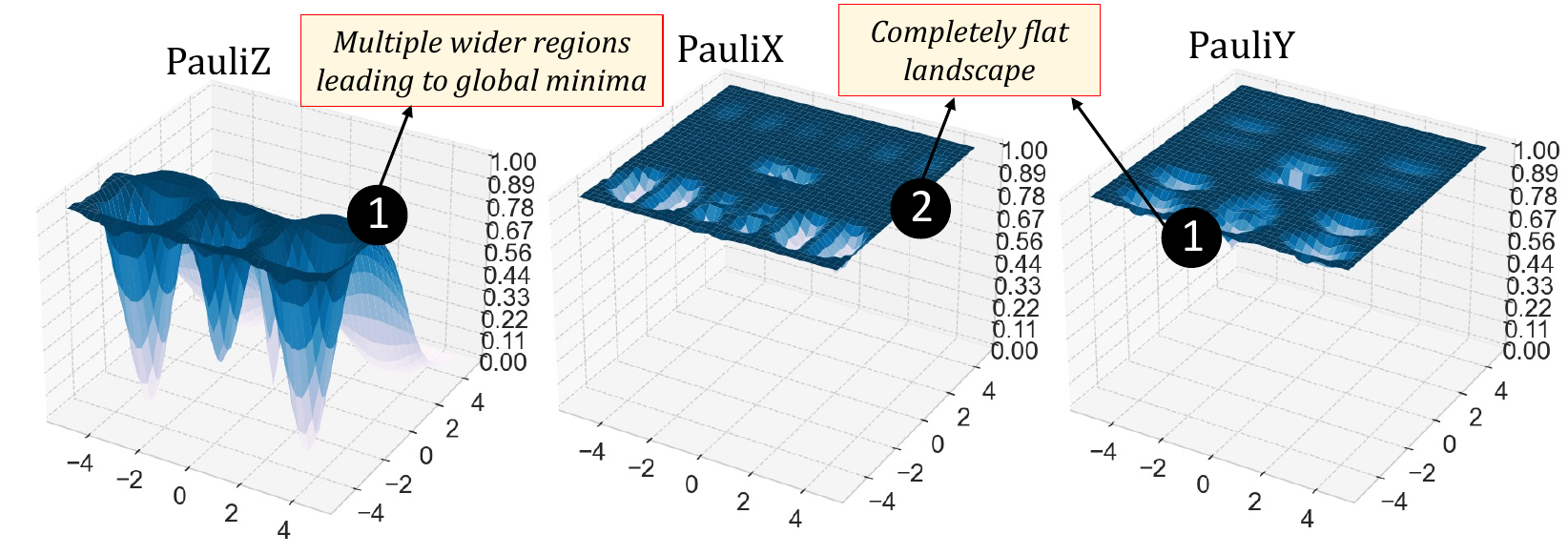}
    \caption{\footnotesize Optimization landscape of 10-Qubit Local QNN with different observables in Noisy Environemnt. The landscapes with PauliZ observable in local QNN still have multiple wider regions leading to solution. PauliX and PauliY observables' landscape are completely flat. }
    \label{fig:10Qubit_localQNN_LS}
\end{figure}

The optimization landscapes in case of 10-qubit local QNN with local cost function design, as shown in Fig. \ref{fig:10Qubit_localQNN_LS}, do not differ much to that of 8-qubit QNN (with local cost function) design (Fig. \ref{fig:8qubit_localQNN_LS}). 
The PauliZ observable still exhibits a pretty dynamic landscape with multiple regions containing the solution making it suitable for the optimizers to find the solution, as shown in label \circled{1} of Fig. \ref{fig:10Qubit_localQNN_LS}. This shows a greater potential of PauliZ observable when used local cost function definition, to overcome the so called BPs problem even at higher qubit count and under noisy conditions.  
On the other hand, the optimization landscape for PauliX and PauliY observables are completely flat and are not suitable for the optimization (label \circled{2} and \circled{3} in Fig. \ref{fig:10Qubit_localQNN_LS}) analogous to the case of 8 qubit QNN with local cost function (\circled{2} and \circled{3} in Fig. \ref{fig:8qubit_localQNN_LS}).

The 10-qubit QNN with local cost function are then trained to learn the problem defined in Equation \ref{eq:CF}. The training results are typically aligned with their optimization landscape where the PauliZ observable outperforms PauliX and PauliY observables, as shown in Fig. \ref{fig:8_10Q_training_local}(right).

\section{Conclusion}
This paper explores the impact of quantum noise on QNN trainability, highlighting how observable selection and cost function design influence the onset of barren plateaus (BPs) in the NISQ era. We show that PauliZ with a global cost function maintains good trainability up to $6–8$ qubits but fails at $10$, while PauliX and PauliY lead to earlier BPs. Our key finding is that a custom Hermitian observable with global cost not only resists noise-induced degradation but leverages noise to enhance training up to $10$ qubits. PauliZ also performs best under local cost settings. Future work will extend to other noise types and broader problem domains.

\end{spacing}

\vspace{-5pt}
\section*{Acknowledgements}
This work was supported in part by the NYUAD Center for Quantum and Topological Systems (CQTS), funded by Tamkeen under the NYUAD Research Institute grant CG008, and the Center for Cyber Security (CCS), funded by Tamkeen under the NYUAD Research Institute Award G1104.

\bibliographystyle{IEEEtran}
\bibliography{main.bib}

\newpage



\vfill

\end{document}